\documentclass[a4paper,11pt]{article}
\pdfoutput=1 

\usepackage{jinstpub} 

\usepackage{siunitx}
\usepackage{booktabs}
\usepackage{subcaption}
\usepackage{graphicx}
\usepackage[symbol]{footmisc}
\usepackage{forest}

\newcommand{\daqname}{CuboDAQ}
\title{The CuboDAQ Data Acquisition System}

\author[a,*,\dag]{E.~Zaffaroni,\note{Corresponding author.}\note{Now at Université de Genève.}}
\author[a]{and G.~Haefeli}

\affiliation[a]{Institute of Physics, Ecole Polytechnique Fédérale de Lausanne (EPFL),\\Lausanne, Switzerland}

\emailAdd{ettore.zaffaroni@cern.ch}

\abstract{
CuboDAQ is a custom data acquisition system to read out SiPM-based detectors.
It features electronic boards to digitize the SiPMs signal, an FPGA-based system-on-module board, the connectivity to transmit the data to a central server, and all the software necessary to operate them.
The front-end is based on the TOFPET2 ASIC, produced by PETsys, connected to a custom board, featuring an Enclustra Mercury SA1 module with a Cyclone~V FPGA.
Multiple boards can be operated synchronously by distributing the clock and synchronous reset signals from a central source.
The system features a complete software framework to calibrate and monitor the detectors, to acquire and process data and to perform track reconstruction.
The CuboDAQ system performance has been evaluated in different scenarios and can cope with sustained rates of above \SI{4e6}{hits/s}, with peaks of more than \SI{7e6}{hits/s}.
This system has been employed for the readout of the SND@LHC detector at CERN, a testbeam telescope and lab setups.
}


\arxivnumber{2410.00190} 

\usepackage{lineno}

\begin{document}
\maketitle
\flushbottom

\section{Introduction}
The \daqname{} is a Data Acquisition (DAQ) system developed to read out detectors based on silicon photomultipliers (SiPMs).
Its primary application is to read out high-density SiPM arrays used in scintillating fibre (SciFi) trackers: a single DAQ board can read out 512 channels in total.
Configurations with fewer channels and different SiPM types are also possible.

CuboDAQ is designed to be scalable, allowing it to be used in laboratory setups, testbeams and in medium-sized experiments.
It is used in all the subsystems of the SND@LHC experiment~\cite{snd_det_paper}, a neutrino detector at the Large Hadron Collider that has been operating since the beginning of Run3 in 2022.
More than \num{16000} channels are present in this system. 
Additionally, it is used for several projects at EPFL\footnote{Ecole Polytechnique Fédérale de Lausanne}: in a testbeam telescope, for the R\&D of new generation SiPMs and for an electromagnetic calorimeter.

The \daqname{} system is composed of hardware and software components.
The custom hardware consists of two electronic boards: a DAQ board, hosting a System-on-Module (SoM) FPGA, and a front-end (FE) board.
The off-the-shelf hardware components are a TTC system~\cite{ttc,ttc_url} hosted in a VME crate, a computer, an Ethernet switch and the power supplies for the boards and SiPMs.
The software consists of the necessary elements to control and calibrate the system and acquire the data.

The FE board is based on the TOFPET2C ASICs~\cite{tofpet2_url,Schug:2018klm,Nadig:2019rjw}, produced by PETsys~\cite{petsys} and can read-out a total of 128 channels (64 per ASIC).
The digitized data from the FE boards is collected by a Cyclone V FPGA and transferred to the on-chip ARM processor, which takes care of transmitting it to a central server.
A total of 4 FE boards can be connected to a single DAQ board.

Multiple DAQ boards are synchronised using the Trigger, Timing and Control (TTC) system~\cite{ttc,ttc_url}, which distributes synchronous clock, reset and trigger commands.

CuboDAQ is intended to be a self-contained data acquisition system built around the TOFPET2C ASIC, but it can also serve as a framework to operate different front-ends: its modular structure enables the integration of different front-end ASICs with the development of only a relatively simple FE board and necessary adaptations to the software and firmware.

The structure of the \daqname{} system and its hardware and software components and the relations between them are described below.
The performance of the system, the operation modes, the data format and the most relevant design choices are also discussed.

\section{Overview}
\begin{figure}[h]
  \centering
  \includegraphics[width=0.95\textwidth]{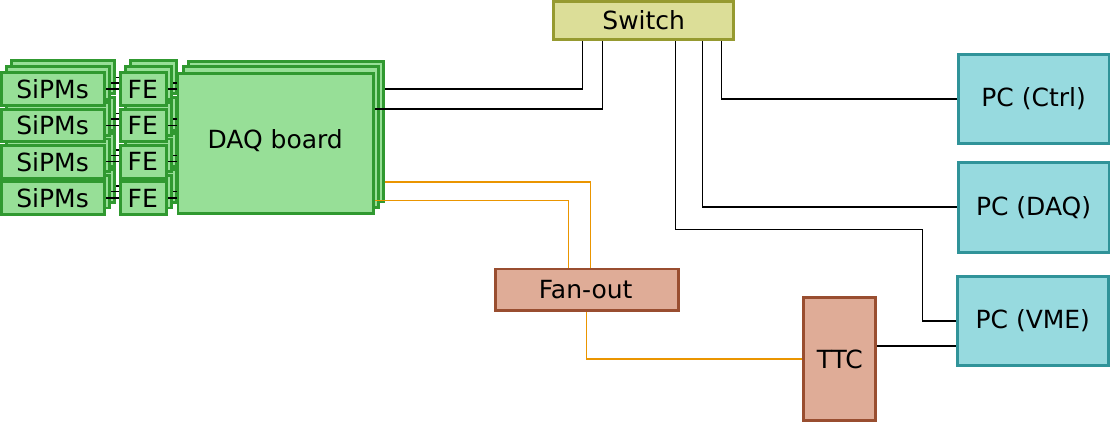}
  \caption{Scheme of the hardware components of the \daqname{}. Computers are drawn in cyan (and identified with PC), the TTC system is in red, the DAQ components in green and the ethernet switch in yellow. Orange lines represent optical fibres, black lines represent copper connections. The power supplies are not shown.}
  \label{fig:hw-scheme}
\end{figure}

The structure of the hardware part of the \daqname{} system, in its most general configuration, is presented in Figure~\ref{fig:hw-scheme}.

\emph{PC (Ctrl)} hosts the software used to control and monitor the full system, and communicate with the other components (other computers and DAQ boards) via TCP/IP.
The DAQ boards collect the data from the front-ends and transmit it to \emph{PC (DAQ)} after minimal processing.
This, in turn, processes the data (event building, noise filtering, etc.) and saves it to disk.
\emph{PC (VME)} is used to control the TTC system, which is hosted in a VME crate.
For simplicity, the various software components can be operated on the same computer.

The TTC system, composed of a TTCvi and a TTCex modules~\cite{ttc,ttc_url}, produces a \SI{40.079}{MHz} clock\footnote{In the rest of the manuscript it will be referred to as ``\SI{40}{MHz} clock'' for simplicity.} and a synchronous reset signal that is delivered to the DAQ boards with optical fibres. 
It can also produce a trigger signal, used either for synchronization verification or for event building, depending on the operation mode (see Section~\ref{sec:sw}).
If a single DAQ board is used, the TTC system is not necessary, as each board can generate its own clock.

\begin{figure}[h]
  \centering
  \includegraphics[width=0.95\textwidth]{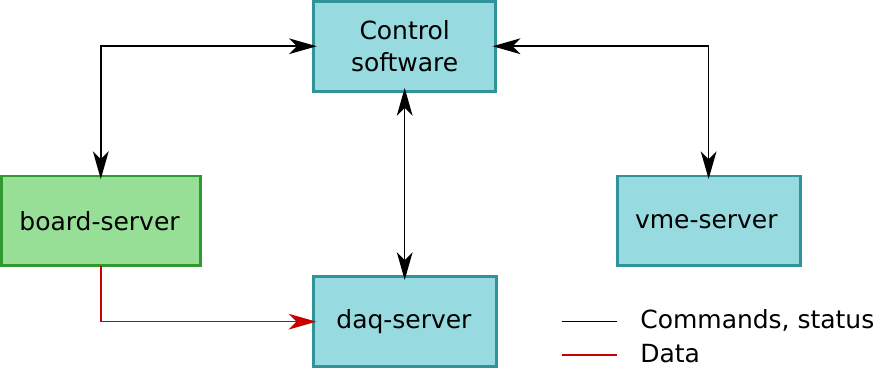}
  \caption{Scheme of the software components of the \daqname{}. The parts in cyan run on standard computers, the part in green on the embedded system in the FPGA.}
  \label{fig:sw-scheme}
\end{figure}

The software components are shown in Figure~\ref{fig:sw-scheme}.
The servers are written in C++, while the control software is based on a Python module.
The software is available in~\cite{cubodaq_sw}.
The \texttt{board-server} executable runs on the ARM processor present on the DAQ board and its main function is to read data from the FPGA and transmit it over the network to the \texttt{daq-server} executable.
Additionally, upon request of the control software, it performs the calibration of the FE ASICs and transmits the status of the board.

The \texttt{daq-server} receives the data from the DAQ boards and processes it before writing it to disk, according to the setting sent by the control software.
The monitoring is also managed by the control software: the \texttt{daq-server} will transmit its status (data received, buffer lengths, etc.) upon request.

The \texttt{vme-server} executable is used to control the TTC system, providing the capability to send a synchronous reset to all the boards at the beginning of the data acquisition, and the trigger signal.

The \emph{Control software} consists of executables based on a dedicated  Python module to operate the various software components.
Example scripts to perform the calibration and run the data acquisition are available in~\cite{cubodaq_example}.

The electronic boards are described in more detail in Section~\ref{sec:boards}, while the TTC system in Section~\ref{sec:ttc}.
These software components are discussed in more details in Section~\ref{sec:sw}.

\section{Electronic boards}
\label{sec:boards}

\subsection{DAQ board and firmware}
\begin{figure}[h]
  \centering
  \includegraphics[width=0.8\textwidth]{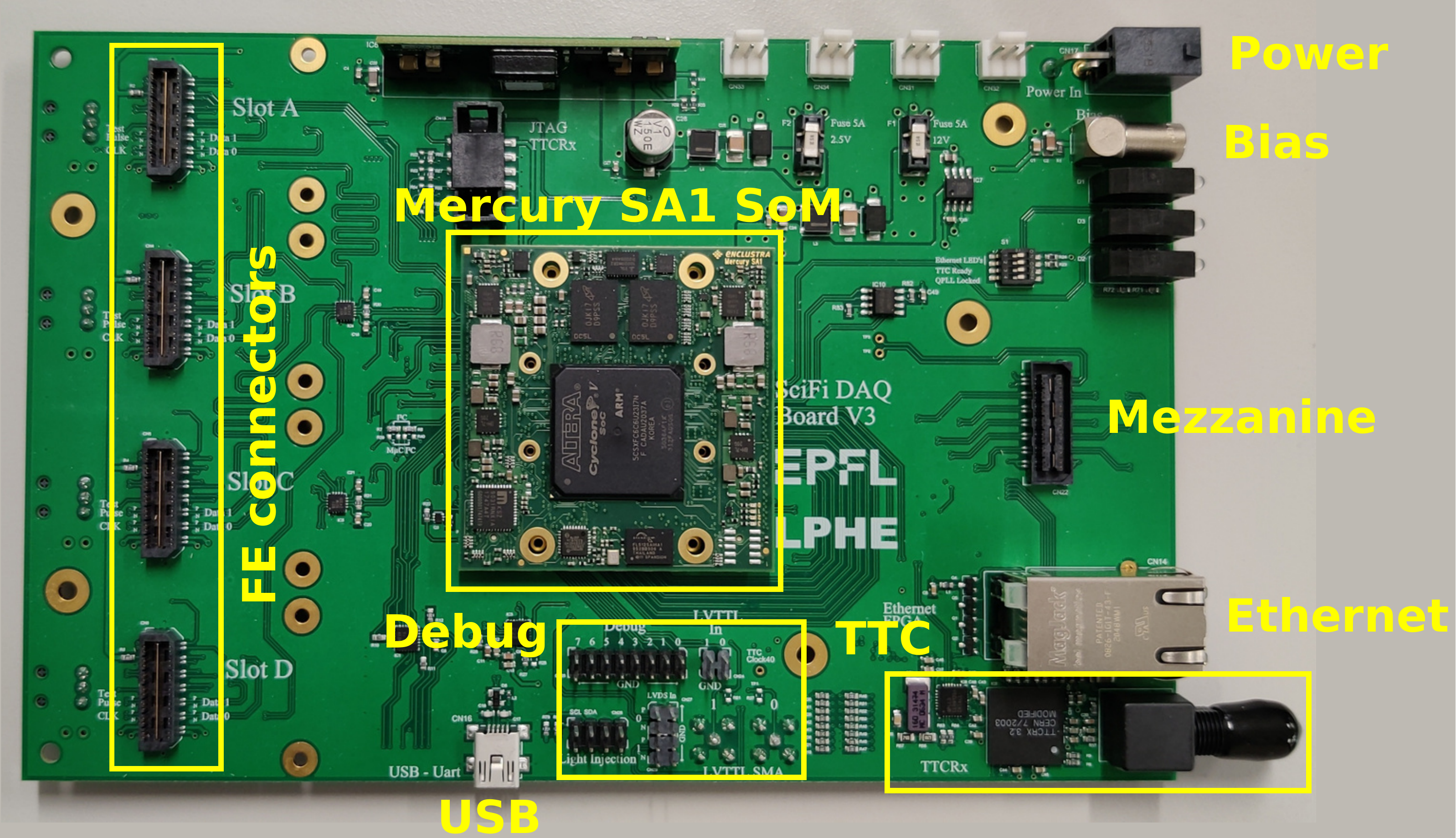}
  \caption{A picture of the DAQ board.} 
  \label{fig:daq-board}
\end{figure}

The DAQ board, shown in Figure~\ref{fig:daq-board}, is based on a Mercury SA1 SoM by Enclustra~\cite{enclustra}, featuring a Cyclone~V SoC FPGA.
The FPGA provides the connectivity to the the FE ASICs for configuration and data transfer.
It also features an on-board ARM processor running a custom Linux distribution provided by Enclustra, to handle the communication to the DAQ server and the control system.
The choice of a commercial SoM greatly simplifies the DAQ board design, as peripherals such as RAM, eMMC, and Ethernet are already implemented, as well as the software development, since an OS compatible with the SoM is already provided.

The communication between the processor and the FPGA is implemented with a Xillybus~\cite{xillybus} IP core, a DMA-based data transport solution between the FPGA and the Linux OS.
The FPGA logic connects to the IP core through standard FIFOs, while the application running on the processor performs standard file IO operations on pipe-like device files.
An FPGA to CPU stream is used for data transfer.
A bidirectional addressed stream, which shows up in the file system as a seekable device file, is used to read and write the FPGA registers.

A DAQ board can host up to four FE boards, described in Section~\ref{subsec:fe}, for a total of up to 512 SiPM channels.
It is powered with \SI{12}{V}, \SI{2.5}{A} from a dedicated connector and provides power to the FE boards, and the bias voltage for the SiPMs, delivered through the FE boards.

\begin{figure}[h]
  \centering
  \includegraphics[width=0.8\textwidth]{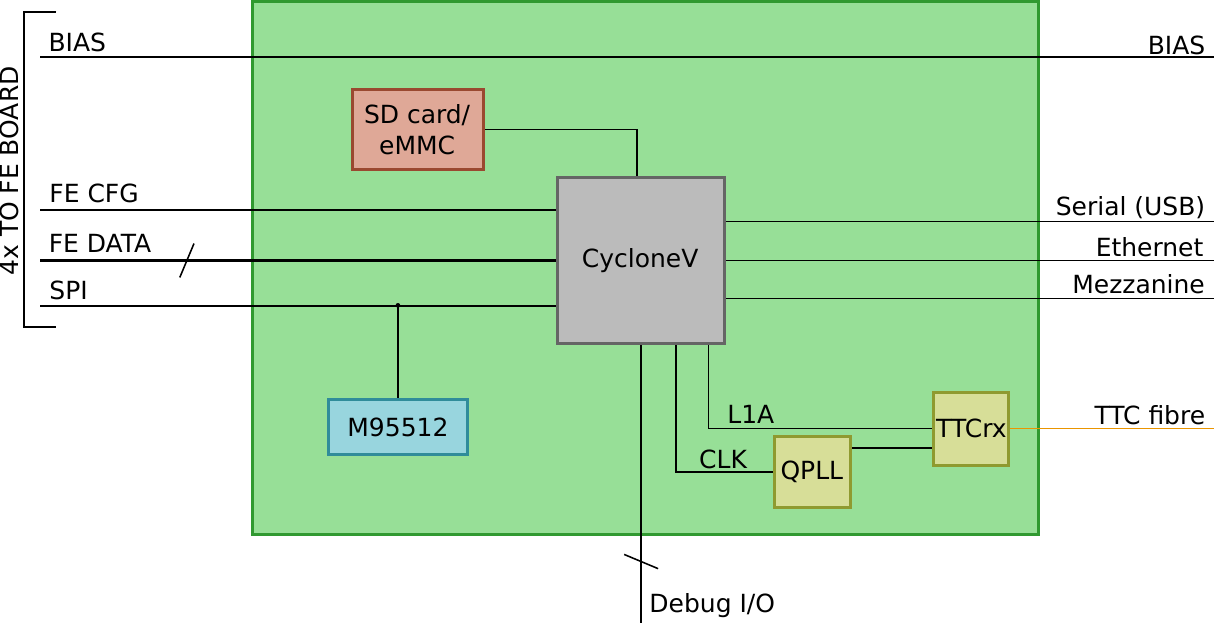}
  \caption{Simplified DAQ board schematic.} 
  \label{fig:daq-board-scheme}
\end{figure}

The relations between the main components of the board are shown in Figure~\ref{fig:daq-board-scheme} and explained below.

The data transmission and control happens through a \SI{1}{Gb/s} Ethernet port.
The clock and synchronous commands can be received from the TTC system through an optical fibre.
These signals are decoded by the TTCrx ASIC, with the clock also being fed through the QPLL ASIC~\cite{ttc,ttc_url}.

A serial terminal is available through the USB connector, enabling the board to be utilized even without a network connection, though with some limitations.
The firmware and operating system is stored on a micro SD card or on the eMMC memory present on the SoM (the latter option is available since revision 4 of the DAQ board).

A multi-pin connector is available to extend the functionality of the board with custom mezzanine boards.
An EEPROM is present on the board to store information like the board ID and the revision.
Debug connectors are also available: several internal FPGA signals can be routed to these with dedicated commands.

\begin{figure}[h]
  \centering
  \includegraphics[width=\textwidth]{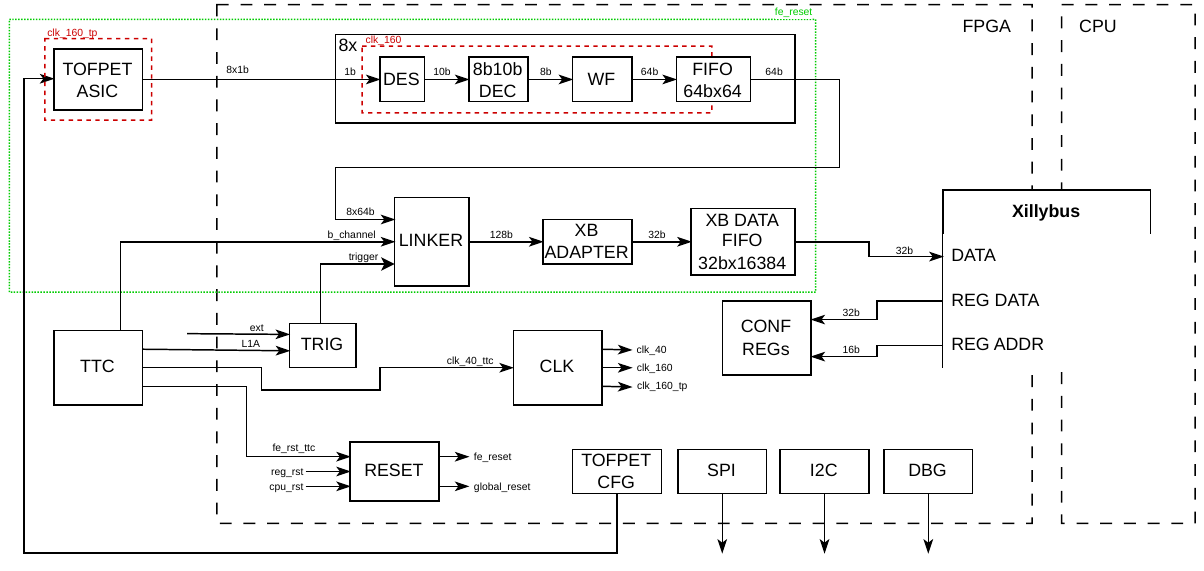}
  \caption{Simplified scheme of the FPGA firmware. The clock domains different from \texttt{clk\_40} are enclosed in the red dashed lines. The \texttt{fe\_reset} domain is enclosed in the green dotted line. The size in bits of each step of the data flow is also shown.} 
  \label{fig:fw-scheme}
\end{figure}

A scheme of the firmware is shown in Figure~\ref{fig:fw-scheme}.
The data is received from each TOFPET2 ASIC a single serial line, in 8b10b encoding.
It is first deserialized (DES), decoded (8b10b DEC) and fed to the word framer (WF), which assembles the received packets in 64-bit words.
These are then transferred in a FIFO to change clock domain from \SI{160}{MHz} to \SI{40}{MHz}.

The TOFPET data, along with the triggers and B-Channel data are assembled in 128-bit packets, and their timestamp is extended to 64 bits (LINKER).
The 128-bit packets are then split in four 32-bit words (XB ADAPTER) before being transferred in another FIFO, from which they are read by the Xillybus IP core and made available on the CPU via the device file \texttt{/dev/xillybus\_data}.

The triggers can be received via the TTC system, from other external sources, or generated with an internal sequencer (TRIG).
The long-format B-Channel data (b\_channel), explained in Section~\ref{sec:ttc}, consists of 32-bit packets that can be transmitted via the TTC system and optionally added in the data stream (not used in normal data taking operations).

The configuration registers (CONF REGS) of the FPGA are mapped to the seekable device file \texttt{/dev/xillybus\_registers}: the address of a register is selected by seeking the desired location in the file before accessing it.

There are three clock domains generated in the FPGA (CLK): two \SI{160}{MHz} ones, one for the data receiver, one for the TOFPET2 ASICs, and a \SI{40}{MHz} one.
They are all generated with Phase-Locked Loops (PLLs) to be all in phase with the clock received via the TTC system or, failing this, a clock generated in the FPGA.
The relative phase of the \SI{160}{MHz} clocks can be adjusted to compensate for cable length, as the FE cards described below can be either connected directly to the DAQ board or through an extension cable.

Two different resets signals can be generated from different sources (RESET).
The \texttt{fe\_reset} only affects the receiver chain and the TOFPETs data transmission and digitization, leaving the FPGA and ASIC configurations unaffected.
This is used at the beginning of data taking, after configuring all the elements, and can be triggered by the TTC system with the short B-Channel command \texttt{0x04}, to synchronize multiple DAQ boards, or with a register write.

The \texttt{global\_reset} resets every element of the FPGA, the TOFPET2s, the TTCrx and the QPLL.
This can be triggered with a register write or is asserted from the ARM processor when booting.

The TOFPET2 ASICs are configured by writing the configuration word in dedicated registers and trasmitting it with a dedicated core (TOFPET CFG).
The FPGA also contains SPI and I$^2$C cores and a debug multiplexer (DBG) that can connect different internal signal to the connectors present on the board.

\subsection{FE board and TOFPET2 ASIC}
\label{subsec:fe}
\begin{figure}[h]
  \centering
  \includegraphics[width=0.5\textwidth]{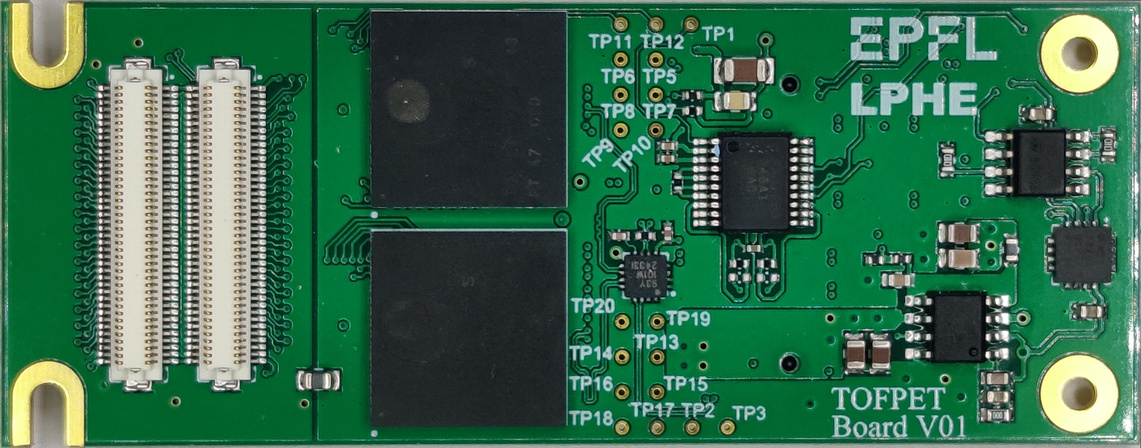}
  \caption{A picture of the FE board. From left to right, the two SiPM connectors, the two TOFPET2C ASICs and the additional chips described in the text are visible. The connector to the DAQ board is on the back.}
  \label{fig:fe-board}
\end{figure}

\begin{figure}[h]
  \centering
  \includegraphics[width=0.8\textwidth]{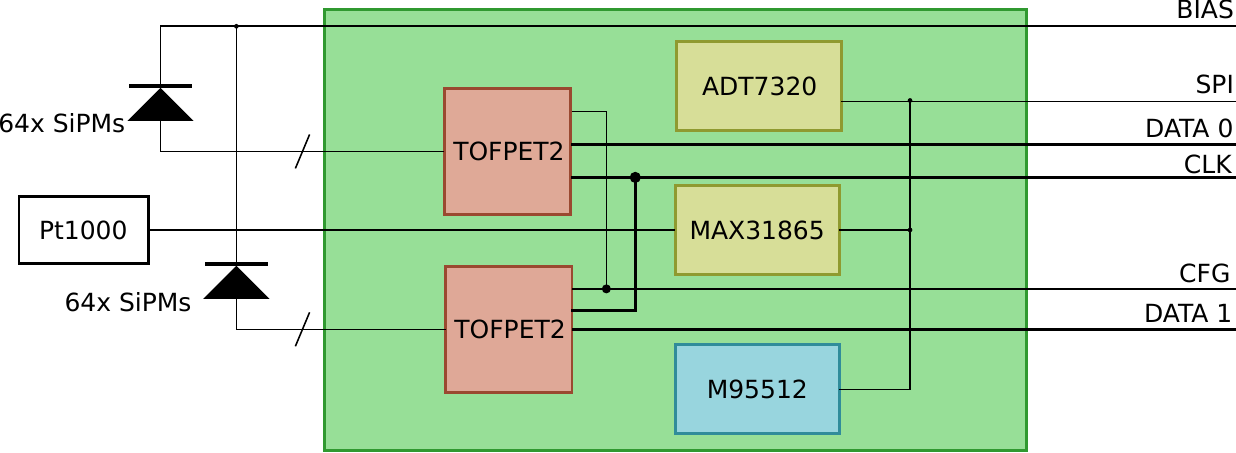}
  \caption{Simplified FE board schematics.}
  \label{fig:fe-scheme}
\end{figure}

The FE board, shown in Figure~\ref{fig:fe-board}, hosts two TOFPET2~\cite{tofpet2_url,Schug:2018klm,Nadig:2019rjw} ASICs, used to digitize the SiPMs signal.
These are connected to the SiPMs via two 80-pin fine pitch connectors and to the DAQ board via a 28-pin connector (not visible).
Additionally, as shown in the simplified schematic of Figure~\ref{fig:fe-scheme}, this board contains a sensor to measure its own temperature (the ADT7320), an IC to measure the temperature of the SiPM with a PT1000 thermistor (MAX31865) and an EEPROM that stores the ID and version of the board (M95512).

The temperature sensors and the EEPROM are configured and read-out using the SPI bus.
The TOFPET2 ASICs use a \SI{160}{MHz} clock, generated with a PLL on the FPGA to be in phase with the \SI{40}{MHz} clock received via the TTCrx.
The ASICs are configured with a dedicated bus, based on SPI, but incompatible with this standard.
The digitized data is transmitted asynchronously at a rate of \SI{160}{Mbit/s} on one LVDS line per ASIC, using 8b10b encoding.

The TOFPET2, revision C, has been chosen as the front-end ASIC for this DAQ system for several reasons: it features 64 channels (compared to the 32 channels of most read-out ASICs), which enables a high channel density while keeping the price per channel relatively low, at around 4.00 -- 4.50~CHF.
It can measure both time and signal amplitude and it does not require an external trigger.
Nonetheless, it has some drawbacks: there is no possibility to trigger externally if needed (this has been solved by implementing the trigger in software, as explained in Section~\ref{sec:daq-server}) and there is very limited sensitivity to the signal amplitude for small SiPMs, like the ones used for the SciFi tracker.

\begin{figure}[h]
  \centering
  \includegraphics[width=\textwidth]{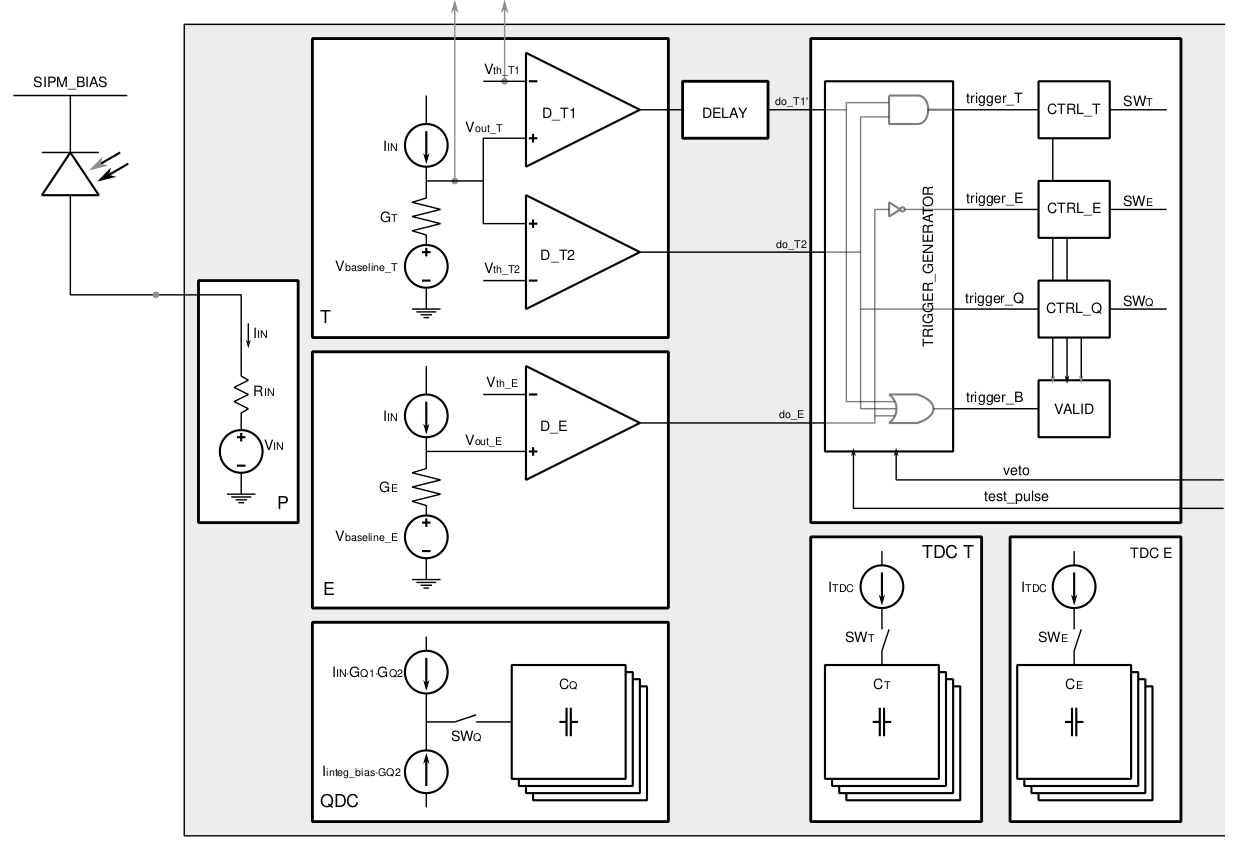}
  \caption{Simplified schematics of a TOFPET2C channel, as presented in its datasheet~\cite{tofpet2_datasheet}. The current from the SiPM is amplified and replicated in the three branches (T, E and QDC). The output of the discriminators T1, T2 and E are combined to generate trigger signals to initiate digitization in the TDC and QDC branches.}
  \label{fig:channel-scheme}
\end{figure}

The TOFPET2 ASIC features 64 independent channels, each containing a low impedance pre-amplifier and two transimpedance amplifiers (TIAs), optimized for time resolution and charge measurement.
The TIAs are identified as T and E: T has a higher gain, E has a larger dynamic range.
The schematics of a channel is shown in Figure~\ref{fig:channel-scheme}.

Up to three discriminators are used to digitize a pulse, enabling the rejection of low-amplitude pulses and ensuring excellent time resolution with minimal channel dead time.
Two Time to Digital Converters (TDCs) with a binning of \SI{40}{ps} and one Charge to Digital Converter (QDC) with a dynamic range of \SI{1500}{pC} are available in each channel.
The signal amplitude can be measured either with Time-over-Threshold (ToT), which exploits both TDCs, or using the QDC.

The three discriminators, identified as T1, T2 and E, have a different dynamic ranges and serve different purposes.
\begin{description}
  \item[T1] is connected to the T TIA and has the lowest dynamic range, so it can be precisely set close to the noise baseline. This discriminator is normally used to measure the timestamp of the hit.
  \item[T2] is connected to the T TIA and has a larger dynamic range. It is used to start the charge integration (if selected) and to validate a hit, if the signals are contained in the dynamic range of the T TIA.
  \item[E] is connected to the E TIA and is used to measure the ToT (if selected) and to validate a hit. It is used if the signal is not contained in the dynamic range of the T TIA.
\end{description}

\noindent The outputs of these discriminators are used to generate 4 trigger signals:

\begin{description}
  \item[trig\_T] The rising edge of trig\_T is used to determine the timestamp of the hit using one of the available TDCs.
  \item[trig\_Q] The rising edge of trig\_Q starts the charge integration, if the channel is used in QDC mode. Otherwise it is only used to validate the hit.
  \item[trig\_E] A rising edge of trig\_E corresponds to the falling edge of the signal and as it is used to determine the pulse duration if the channel is in ToT mode. Otherwise it is only used to validate the hit.
  \item[trig\_B] This signal is high if any of the discriminators output is high. On the falling edge of trig\_B, the hit is validated: if a rising edge was detected on the three other triggers, the hit is accepted and queued for transmission, otherwise it is rejected. 
\end{description}

\noindent The details regarding the different discriminator settings combinations can be found in the TOFPET2 datasheet~\cite{tofpet2_datasheet}.

\section{Trigger, Timing and Control system}
\label{sec:ttc}
The Trigger, Timing and Control (TTC) system has been developed at CERN and is used to transmit the clock, trigger and synchronous signals to the electronic controllers of the LHC experiments~\cite{ttc}.
It is a unidirectional optical fibre based transmission system, where two information channels, A and B, are multiplexed and encoded using the LHC Bunch Crossing clock as the carrier frequency~\cite{ttc_modules}.

The A-Channel carries exclusively the \emph{Level-1 Trigger Accept} (L1A) information while the B-Channel carries packaged address and data information for the sending of various reset commands or calibration, control and test parameters.
The data packages sent on the B-Channel can either be of short format (8 data bits), used for broadcast commands or of long format (32 total bits) for individually addressed commands or data transfers~\cite{ttc_modules}. 

In the context of \daqname{}, it is primarily used to transmit the clock, trigger and synchronous reset signals to all the DAQ boards used in a detector.
The TTC setup features a TTCvi and a TTCex modules hosted in a VME crate, identified as TTC in Figure~\ref{fig:hw-scheme}.
The TTCvi produces the signals to be transmitted to each board, while the TTCex encodes them and broadcasts them from a single laser source to several destinations over a passive network composed of a hierarchy of optical tree couplers.

Each DAQ board hosts an optical fibre receiver together with a TTCrx and a QPLL ASICs.
The former decodes the signal received from the optical fibre, recovers the clock, the trigger pulses and synchronous commands, while the latter is used to reduce the jitter on the clock using a PLL, in combination with a second PLL in the FPGA.

The clock can be generated by the TTCex module or received externally.
The clock must have a precise frequency, in the range between \SI{40.0749}{MHz} and \SI{40.0823}{MHz}, to allow the QPLL to lock.
In revision 4 of the DAQ board, the QPLL can be bypassed to offer greater tolerance on the clock frequency.
No measurable effect on the time resolution has been observed in this case.

\section{Software}
\label{sec:sw}

The software is composed of several executables written in C++, running on the DAQ boards and on the computers used to acquire data, and a Python module, used for controlling the different parts of the system.
The software is available in~\cite{cubodaq_sw}.

\subsection{DAQ board server}
Each DAQ board features a Cyclone~V SoC FPGA, running a custom Linux distribution.
The main executable running on it is the \texttt{board-server}, whose main purpose is to receive and execute commands from the control system, read the data from the FPGA and transmit it to the DAQ server.

The main functionalities of the board server include:
\begin{itemize}
  \setlength\itemsep{0em}
  \item calibrate and configure the front-ends;
  \item configure the FPGA;
  \item obtain the status of the FPGA (FIFOs, counters, etc.)~and of the board (temperature of front-ends and SiPMs) and transmit it to the control system;
  \item start and stop the data acquisition;
  \item enable debug functionalities;
\end{itemize}
For details about the commands and the options, refer to the documentation in the software repository~\cite{cubodaq_doc}.

\subsection{DAQ server}
\label{sec:daq-server}
The DAQ server receives data from the DAQ boards in the form of hit and trigger words, performs event building, event processing and writes the data to disk.

The event builder receives the hits from all the boards, and sorts them by timestamp.
The event building algorithm can be chosen between a triggerless one, i.e.~hits are grouped into events purely based on their timestamp, or a triggered one, where hits are grouped only in presence of a trigger word in the data stream.
The system does not feature a hardware trigger, in both modes all the hits recorded by the DAQ boards are transmitted to the DAQ server.

When running in triggerless mode, the time window during which hits are grouped into the same event can be selected.
When running in triggered mode, the trigger latency and trigger window can be selected.

Events can be further processed by different types of processors such as calibrators, to perform online calibration of the data, or filters, to reject unwanted events (normally used in triggerless mode).
After the last processing step, events are written to disk.

\subsection{TTC VME control}
The \texttt{vme-server} executable controls the TTC system hosted in a VME crate.
It runs on a standard computer and can communicate to the VME crate via the Wiener~\cite{wiener} VM-USB, or the CAEN~\cite{caen} Vx718 VME interfaces.

The \texttt{vme-server} receives commands from the control system and relays them to the TTCvi.
It configures the TTCvi as needed to produce a synchronous reset command, in order to have the timestamps recorded by the DAQ boards completely synchronized.
Additionally, it configures the TTCvi to transmit the relevant trigger depending on the selected event builder.

The trigger signal, called \emph{Level-1 Accept} or \emph{L1A} in the TTC context, is used as a periodic heartbeat when acquiring data in triggerless mode and as trigger when running in triggered mode.
This can be generated internally in the TTCvi or received from an external source as a NIM or ECL signal.

The synchronous reset consists of a short B-Channel command, that can optionally be synchronised to an external pulse connected to one of the B-Go inputs of the TTCvi, e.g.~in SND@LHC the reset is synchronised to the orbit signal of the LHC~\cite{snd_det_paper}.

\subsection{Control system}
The control system is based on a Python package, containing the classes to control and monitor all parts of the \daqname{} system.
The most relevant classes are:
\begin{itemize}
  \setlength\itemsep{0em}
  \item \texttt{DaqBoards}, used to calibrate, control and monitor multiple DAQ boards;
  \item \texttt{DaqServerControl}, used to control and monitor the \texttt{daq-server};
  \item \texttt{VmeClient}, used to control the \texttt{vme-server};
  \item \texttt{Daq}, a wrapper of the three classes listed above to manage the most common data taking situations.
\end{itemize}

These classes allow the user to build their own data acquisition framework, customized to the requirements of their specific application.
Some example calibration and data taking scripts can be found in~\cite{cubodaq_example}.

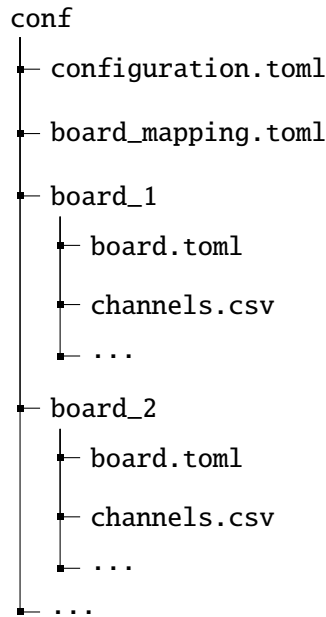
\begin{figure}[h]
  \centering
  \begin{forest}
    for tree={
    font=\ttfamily,
    grow'=0,
    child anchor=west,
    parent anchor=south,
    anchor=west,
    calign=first,
    edge path={
      \noexpand\path [draw, \forestoption{edge}]
      (!u.south west) +(7.5pt,0) |- node[fill,inner sep=1.25pt] {} (.child anchor)\forestoption{edge label};
    },
    before typesetting nodes={
      if n=1
        {insert before={[,phantom]}}
        {}
    },
    fit=band,
    before computing xy={l=15pt},
  }
  [conf
    [configuration.toml]
    [board\_mapping.toml]
    [board\_1
      [board.toml]
      [channels.csv]
      [\dots]
    ]
    [board\_2
      [board.toml]
      [channels.csv]
      [\dots]
    ]
    [\dots]
  ]
  \end{forest}
  \caption{Typical configuration folder structure for the \daqname{} software.}
  \label{fig:conf-structure}
\end{figure}

The control software relies on a set of configuration files, structured as shown in Figure~\ref{fig:conf-structure}.

The main folder must contain a \texttt{configuration.toml} and a \texttt{board\_mapping.toml} files.
The former contains the global system configuration, including the IP addresses of the servers, the desired event builder, the configuration for the processors, and a list of the DAQ boards with the path to their configuration folder and their IP address.
The latter contains the mapping from detector planes and subsystems to DAQ board IDs, needed for example for noise filtering or subsequent reconstruction of the saved data.

The main folder must then contain a directory for each DAQ board, with the global board settings in \texttt{board.toml} and the per-channel settings in \texttt{channels.csv}.
The remaining files are generated during calibration and the detailed description of their content can be found in~\cite{cubodaq_doc}.

\section{Data format}
\label{sec:data-format}
The data collected by the DAQ server is divided in runs, each corresponding to a continuous data taking period.
Event timestamps are guaranteed to be accurate down to a single clock cycle within the same run.

The data collected in each run is stored in a single folder, containing the data itself and all the configuration and calibration information.

The data itself is saved in files of the format \texttt{data\_NNNN.root} (or \texttt{data\_NNNN.raw} when saving unprocessed packets, see below).
\texttt{NNNN} is an incremental number starting from 0, enabling the data to be split into several files to limit the size of each one.

The remaining files contain the configuration of the boards and front-ends, and the calibration parameters:
\begin{itemize}
  \item \texttt{boards.csv} contains the global board settings of each DAQ board, retrieved from each \texttt{board.toml} file in the configuration folder.
  \item \texttt{fe.csv} contains the global front-end settings of each  ASIC for all DAQ boards, retrieved from each \texttt{board.toml} file in the configuration folder.
  \item \texttt{channels.csv} contains the channel configuration of all channels in all front-ends, retrieved from each \texttt{channels.csv} file in the configuration folder.
  \item \texttt{configuration.json} contains the system configuration, retrieved from the \texttt{configuration.toml} file.
  \item \texttt{board\_mapping.json} contains the board mapping, retrieved from the \texttt{board\_mapping.toml} file in the configuration folder.
  \item \texttt{tdc\_cal.csv} and \texttt{qdc\_cal.csv} contain the calibration constants for the TDC and QDC of each channel in each front-end. They are necessary to apply the calibration during data taking.
\end{itemize}

\subsection{Event data format}
\label{sec:evt-data-format}
The reconstructed events are saved in ROOT~\cite{BRUN199781} files, in a single \texttt{TTree}.
Each entry of the TTree contains an event, composed of global event information (such as timestamp and flags) and a list of hits, containing hit identification and calibration quality information.

The most important branches are the following (the hit branches are arrays and are identified with the square brackets):
\begin{itemize}
  \item \texttt{evt\_timestamp} contains the event timestamp, starting from the system reset, measured in clock cycles of the \SI{160}{MHz} clock.
  \item \texttt{evt\_number} contains the event number (in triggerless mode it is a simple counter, in triggered mode it corresponds to the trigger number).
  \item \texttt{evt\_flags} contains the event flags, set by the noise filters, to determine which condition accepted the event. User flags can also be specified.
  \item \texttt{n\_hits} contains the number of hits in the event.
  \item \texttt{board\_id[n\_hits]} contains the ID of the board that produced each hit.
  \item \texttt{tofpet\_id[n\_hits]} contains the ID of the TOFPET2 ASIC that produced each hit.
  \item \texttt{tofpet\_channel[n\_hits]} contains the ID of the TOFPET channel that produced each hit.
  \item \texttt{timestamp[n\_hits]} contains the calibrated timestamp of the hit within the event, measured in clock cycles of the \SI{160}{MHz} clock. The global hit timestamp is given by adding \texttt{evt\_timestamp}.
  \item \texttt{value[n\_hits]} contains the calibrated QDC or ToT value of the hit, depending on the channel setting.
\end{itemize}

The remaining branches contain uncalibrated data and calibration quality checks. The full description of these branches is given in~\cite{cubodaq_doc}.

\subsection{Raw data format}
\label{sec:raw-data-format}
Data is transmitted from the DAQ boards to the DAQ server in the form of 128-bit packets, organized in four 32-bit words.

The main data packets used during data taking are the hit and trigger ones.
They correspond to a digitized hit in the front-end and to a trigger received by the board respectively.

Data in this format can also be saved to disk by the DAQ server, by selecting the relevant options in the configuration file.
In this case, no processing occurs and the received data is saved to disk in a binary file.
Each packet will occupy 20 bytes: 4 are used to store the board ID that produced the packet and the remaining 16 will store the four 32-bit words composing the hit or trigger packet.
Data is saved in little-endian format.

\section{Performance}
The performance of the \daqname{} has been evaluated with laboratory measurements, testbeams, and in the SND@LHC detector.

The TOFPET2 ASIC can sustain a maximum hit rate of \SI{\sim 2}{MHz} before being limited by the data link bandwidth.
This limit is rarely reached because other parts of the system will act as a bottleneck.

A single DAQ board can sustain a maximum hit rate of \SI{\sim 1.3}{MHz} from all front-ends, before the data transfer speed becomes a bottleneck.

\begin{figure}[h]
  \centering
  \begin{subfigure}[c]{0.49\textwidth}
      \centering
      \includegraphics[width=\textwidth]{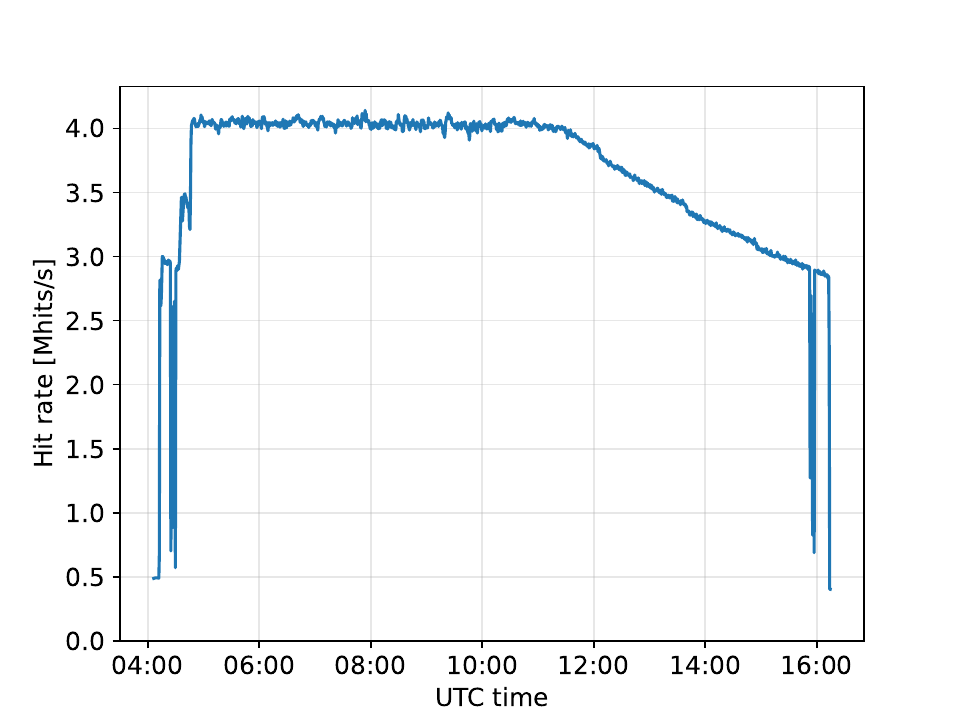}
      \caption{}
      \label{fig:rate-lhc}
  \end{subfigure}
  \begin{subfigure}[c]{0.49\textwidth}
      \centering
      \includegraphics[width=\textwidth]{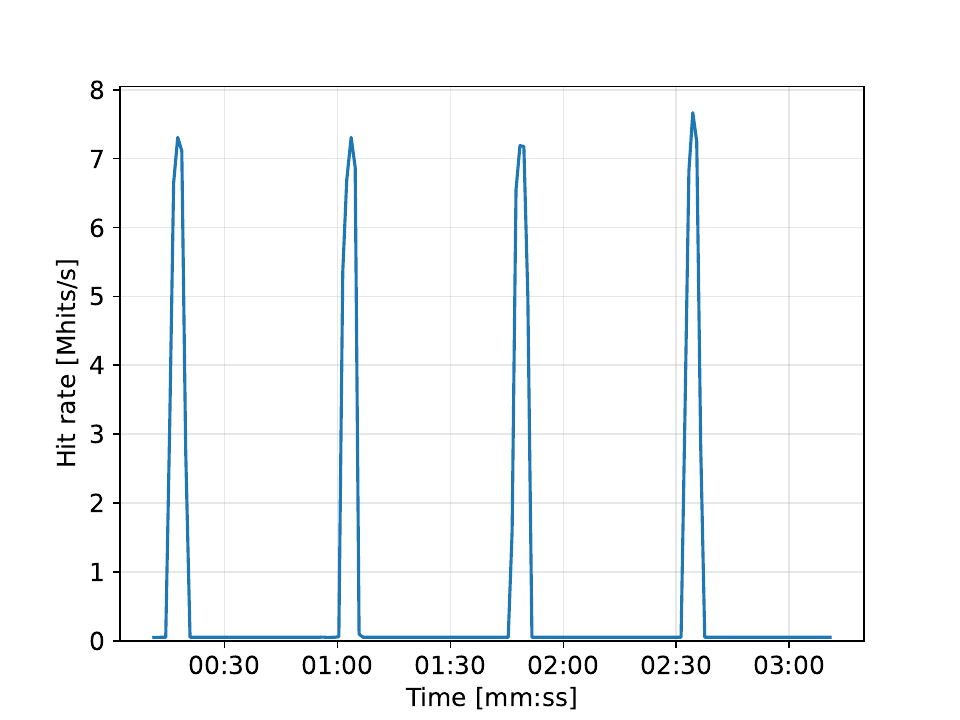}
      \caption{}
      \label{fig:rate-tb}
  \end{subfigure}
     \caption{Left: hit rate recorded by the DAQ system of the SND@LHC experiment during a high-intensity LHC fill. Right: hit rate recorded by the \daqname{} system during a testbeam at CERN SPS.}
     \label{fig:rate}
\end{figure}

The DAQ server running the event builder, assuming it is running on a modern machine with at least 8 processor cores and \SI{32}{GB} of RAM, can process \num{> 4e6} hits per second (from all boards) on average, as seen in Figure~\ref{fig:rate-lhc}.

The instantaneous rate can be an order of magnitude higher, as long as the average rate is acceptable.
This is useful for example at testbeams and in all those conditions where the data is produced in \emph{spills}.
The DAQ has been successfully validated up to an instantaneous rate of \num{> 7e6} hits per second during a testbeam at the Super Proton Synchrotron at CERN, as seen in Figure~\ref{fig:rate-tb}.

The maximum rate of events passing the noise filter and being written to disk depends significantly on the capabilities of the machine running the DAQ server, in particular on the writing speed of the storage device, but also on the size of a single event.

\section{Reconstruction}
\label{sec:reco}
The \daqname{} integrates a reconstruction framework, based on the proteus software~\cite{kiehn_2019_proteus}, designed to reconstruct straight tracks in detectors such as the SND@LHC SciFi tracker, and extrapolate them on the detectors under test.

It reads the data produced by the \texttt{daq-server} and relies on three configuration files describing the detector setup:
\begin{itemize}
  \item \texttt{configuration.toml} contains the configuration of the processing steps (clusterisation, track reconstruction, event filtering) and of the output file.
  \item \texttt{device.toml} contains the description of the detector in terms of sensors used and their type.
  \item \texttt{geometry.toml} contains the position and rotation of each sensor.
\end{itemize}

The software has been developed to be useful in a wide variety of data taking scenarios.
It takes care of clustering (i.e.~grouping adjacent hits into clusters), track fitting, correction of the propagation time of light in the fibres (necessary to reach a ~\SI{100}{ps} time resolution) and matching of the closest cluster on the device under test.

The reconstruction software produces ROOT files containing several branches:
\begin{itemize}
  \item \texttt{events} containing global event information (such as timestamp and counter);
  \item \texttt{tracks} containing global track information (position, slope and quality);
  \item \texttt{hits} containing the hits that compose the clusters (position, timestamp);
  \item \texttt{clusters} containing the clusters (position, timestamp, track association);
  \item \texttt{intercepts} containing the intersections of the tracks with the detector planes (the information in this branch is technically redundant, but it greatly simplifies the data analysis).
\end{itemize}

The data produced by the reconstruction can then be used to study the performance of the detectors under test.
More information about the reconstruction can be found in~\cite{cubodaq_doc} and examples can be found in~\cite{cubodaq_example_reco}.

\section{Conclusions}
The CuboDAQ is a DAQ system for SiPMs based on the TOFPET2 ASIC.
It is composed of two types of electronics boards, one containing the front-end ASICs and one to collect the data from several front-ends and to transmit it to a central server.
Software to operate several boards, collect the data, perform event building and filtering has been developed and is available in~\cite{cubodaq_sw}.
The software also features a track reconstruction framework.

The CuboDAQ system performance has been evaluated in a laboratory environment and during a testbeam campaign, showing a sustained rate capability in excess of \num{\sim 4e6} hits per second, with peaks of \num{> 7e6} hits per second.
The CuboDAQ system is currently being used in the SND@LHC detector since the beginning of its operation in 2022~\cite{snd_det_paper}.

\section*{Acknowledgements}
We thank the electronic and mechanical workshops of the Institute of Physics of EPFL for their support in the development and production of the boards and modules.
We thank Anna Mascellani for her help in the debugging of the calibration routines, and Anni Kauniskangas and Federico Ronchetti for their help in testing the CuboDAQ system and for the careful reading of this document.



\bibliographystyle{JHEP} %
\bibliography{references}

\providecommand{\href}[2]{#2}\begingroup\raggedright\begin{thebibliography}{10}

\bibitem{snd_det_paper}
G.~Acampora, C.~Ahdida, R.~Albanese, C.~Albrecht, A.~Alexandrov, M.~Andreini
  et~al., \emph{{SND@LHC}: the scattering and neutrino detector at the {LHC}},
  \href{https://doi.org/10.1088/1748-0221/19/05/P05067}{\emph{Journal of
  Instrumentation} {\bfseries 19} (2024) P05067}.

\bibitem{ttc}
B.~Taylor, \emph{{TTC} distribution for {LHC} detectors},
  \href{https://doi.org/10.1109/23.682644}{\emph{IEEE Transactions on Nuclear
  Science} {\bfseries 45} (1998) 821}.

\bibitem{ttc_url}
{Baron, S. and others}, ``{The TTC Website}.'' \url{https://ttc.web.cern.ch/}.

\bibitem{tofpet2_url}
{PETsys electronics}, ``{PETsys TOFPET2 ASIC}.''
  \url{https://www.petsyselectronics.com/web/public/products/1}, accessed July
  5, 2024.

\bibitem{Schug:2018klm}
D.~Schug, V.~Nadig, B.~Weissler, P.~Gebhardt and V.~Schulz, \emph{{Initial
  Measurements with the PETsys TOFPET2 ASIC Evaluation Kit and a
  Characterization of the ASIC TDC}},
  \href{https://doi.org/10.1109/TRPMS.2018.2884564}{\emph{IEEE Transactions on
  Radiation and Plasma Medical Sciences} {\bfseries 3} (2019) 444}.

\bibitem{Nadig:2019rjw}
V.~Nadig, D.~Schug, B.~Weissler and V.~Schulz, \emph{{Evaluation Of The PETsys
  TOFPET2 ASIC In Multi-Channel Coincidence Experiments}},
  \href{https://doi.org/10.1186/s40658-021-00370-x}{\emph{{EJNMMI Physics}}
  (2021) } [\href{https://arxiv.org/abs/1911.08156}{{\ttfamily 1911.08156}}].

\bibitem{petsys}
{PETsys electronics, Taguspark, Ed. Tecnologia, 3.2 n.61-64, 2740-257 Porto
  Salvo, Portugal}. \url{https://www.petsyselectronics.com/web/}.

\bibitem{cubodaq_sw}
{E. Zaffaroni}, ``{CuboDAQ software}.''
  \url{https://gitlab.cern.ch/cubodaq/cubodaq-sw}.

\bibitem{cubodaq_example}
{E. Zaffaroni}, ``{CuboDAQ example scripts}.''
  \url{https://gitlab.cern.ch/cubodaq/cubodaq-example-daq}.

\bibitem{enclustra}
{Enclustra GmbH, R\"{a}ffelstrasse 28, CH-8045 Zurich}.
  \url{https://www.enclustra.com/en}, accessed July 5, 2024.

\bibitem{xillybus}
{Xillybus Ltd., P.O.B. 7842, Haifa 31078, Israel}. \url{https://xillybus.com/}.

\bibitem{tofpet2_datasheet}
{PETsys electronics}, ``{TOFPET2 ASIC Datasheet}.''
  \url{https://www.petsyselectronics.com/web/website/documentation/TOFPET2%20Downloads/Documentation/PETsys%20TOFPET%202C%20ASIC%20-%20Datasheet%20(rev%2014).pdf},
  accessed July 5, 2024. Requires login.

\bibitem{ttc_modules}
{P. G\"{a}lln\"{o}}, ``{Modules development for the TTC system}.''
  \url{https://cds.cern.ch/record/433828}, 1999.
\newblock 10.5170/CERN-1999-009.327.

\bibitem{cubodaq_doc}
{E. Zaffaroni}, ``{CuboDAQ documentation}.''
  \url{https://gitlab.cern.ch/cubodaq/cubodaq-sw/-/wikis/home}.

\bibitem{wiener}
{W-IE-NE-R, Linde 18, D-51399 Burscheid, Germany}.
  \url{https://www.wiener-d.com/}.

\bibitem{caen}
{CAEN S.p.A., Via Vetraia 11, 55049 Viareggio (LU), Italy}.
  \url{https://www.caen.it/}.

\bibitem{BRUN199781}
R.~Brun and F.~Rademakers, \emph{{ROOT} -- an object oriented data analysis
  framework},
  \href{https://doi.org/https://doi.org/10.1016/S0168-9002(97)00048-X}{\emph{Nuclear
  Instruments and Methods in Physics Research Section A: Accelerators,
  Spectrometers, Detectors and Associated Equipment} {\bfseries 389} (1997)
  81}.

\bibitem{kiehn_2019_proteus}
M.~Kiehn, ``Proteus beam telescope reconstruction.''
  \url{https://doi.org/10.5281/zenodo.2586736}.

\bibitem{cubodaq_example_reco}
{E. Zaffaroni}, ``{CuboDAQ example reconstruction scripts}.''
  \url{https://gitlab.cern.ch/cubodaq/cubodaq-example-reco}.

\end{thebibliography}\endgroup

\end{document}